\pdfoutput = 1
\RequirePackage[2020-02-02]{latexrelease}
\documentclass[aip,jmp,amsmath,amssymb,reprint]{revtex4-1}

\usepackage{graphicx}
\usepackage{dcolumn}
\usepackage[utf8]{inputenc}
\usepackage[T1]{fontenc}

\usepackage{tikz}
\usepackage{tikz-cd}
\usetikzlibrary{tikzmark,calc,arrows,shapes,decorations.pathmorphing,decorations.pathreplacing}
\usetikzlibrary{arrows}
\usepackage{collectbox}

\makeatletter
\newcommand{\sqbox}{%
    \collectbox{%
        \@tempdima=\dimexpr\width-\totalheight\relax
        \ifdim\@tempdima<\z@
            \fbox{\hbox{\hspace{-.5\@tempdima}\BOXCONTENT\hspace{-.5\@tempdima}}}%
        \else
            \ht\collectedbox=\dimexpr\ht\collectedbox+.5\@tempdima\relax
            \dp\collectedbox=\dimexpr\dp\collectedbox+.5\@tempdima\relax
            \fbox{\BOXCONTENT}%
        \fi
    }%
}
\makeatother

\graphicspath{{./figures/}} 

\begin{document}

\title[Quiver Yangians and Crystal Meltings: A Concise Summary]{Quiver Yangians and Crystal Meltings: \\ A Concise Summary}
\author{Masahito Yamazaki}
\email{masahito.yamazaki@ipmu.jp}
\homepage{\\ webpage: \url{http://member.ipmu.jp/masahito.yamazaki/index.shtml}; \\ associated slides: \url{http://member.ipmu.jp/masahito.yamazaki/files/2021/20210802_ICMP.pdf}}
\affiliation{Kavli Institute for the Physics and Mathematics of the Universe (WPI), University of Tokyo, Kashiwa, Chiba 277-8583, Japan}

\affiliation{Trans-Scale Quantum Science Institute, \\ University of Tokyo, Tokyo 113-0033, Japan}

\date{\today}

\begin{abstract}
The goal of this short article is to summarize some of the recent developments in the quiver Yangians and crystal meltings. This article is based on a lecture delivered by the author at International Congress on Mathematical Physics (ICMP), Geneva, 2021.
\end{abstract}

\maketitle


\section{Introduction}

One of the fascinating aspects of integrable models is that they are located at the intersection of many different topics in mathematics and physics. This comment applies very well to the contents of this paper---even within the limited pages, we will encounter branches of mathematics such as geometry, algebra, representation theory, and combinatorics, and those of physics such as gauge theory and string theory.

The goal of this note is to discuss (1) a class of newly-introduced algebras called the (shifted) quiver Yangians, 
and (2) their representations in terms of the statistical-mechanical model of crystal melting.
Our primary task here is to give a leisurely introduction to the main ideas and results in the recent developments.\footnote{See also Ref.\ \onlinecite{Rapcak:2021hdh} for another introduction on related topics.}
This inevitably means that our presentation will omit many important details and our references will be incomplete;
for full treatments readers are encouraged to consult original papers\cite{Li:2020rij,Galakhov:2020vyb,Galakhov:2021xum} on the subject.\footnote{While we concentrate on the rational cases of quiver Yangians,
the story generalizes to trigonometric\cite{Noshita:2021ldl,Galakhov:2021vbo,Noshita:2021dgj} and elliptic\cite{Galakhov:2021vbo} counterparts.}

\section{(Shifted) Quiver Yangians}

\subsection{Quiver and Superpotential}

Yangians\cite{Drinfeld:1985rx, Drinfeld:1986} are often defined from a Lie algebra $\mathfrak{g}$,
which is determined by a (generalized) Cartan matrix. 
We here instead start with a quiver $Q$ and a superpotential $W$.
Here a quiver $Q$ consists of a set of vertices $Q_0$ and that of arrows $Q_1$,
and a superpotential $W$ is a formal linear sum of oriented cycles of the quiver. Physically, this is the defining 
data for an $\mathcal{N}=4$ supersymmetric quiver quantum mechanics.

\begin{figure}[htbp]
\includegraphics[scale=0.45]{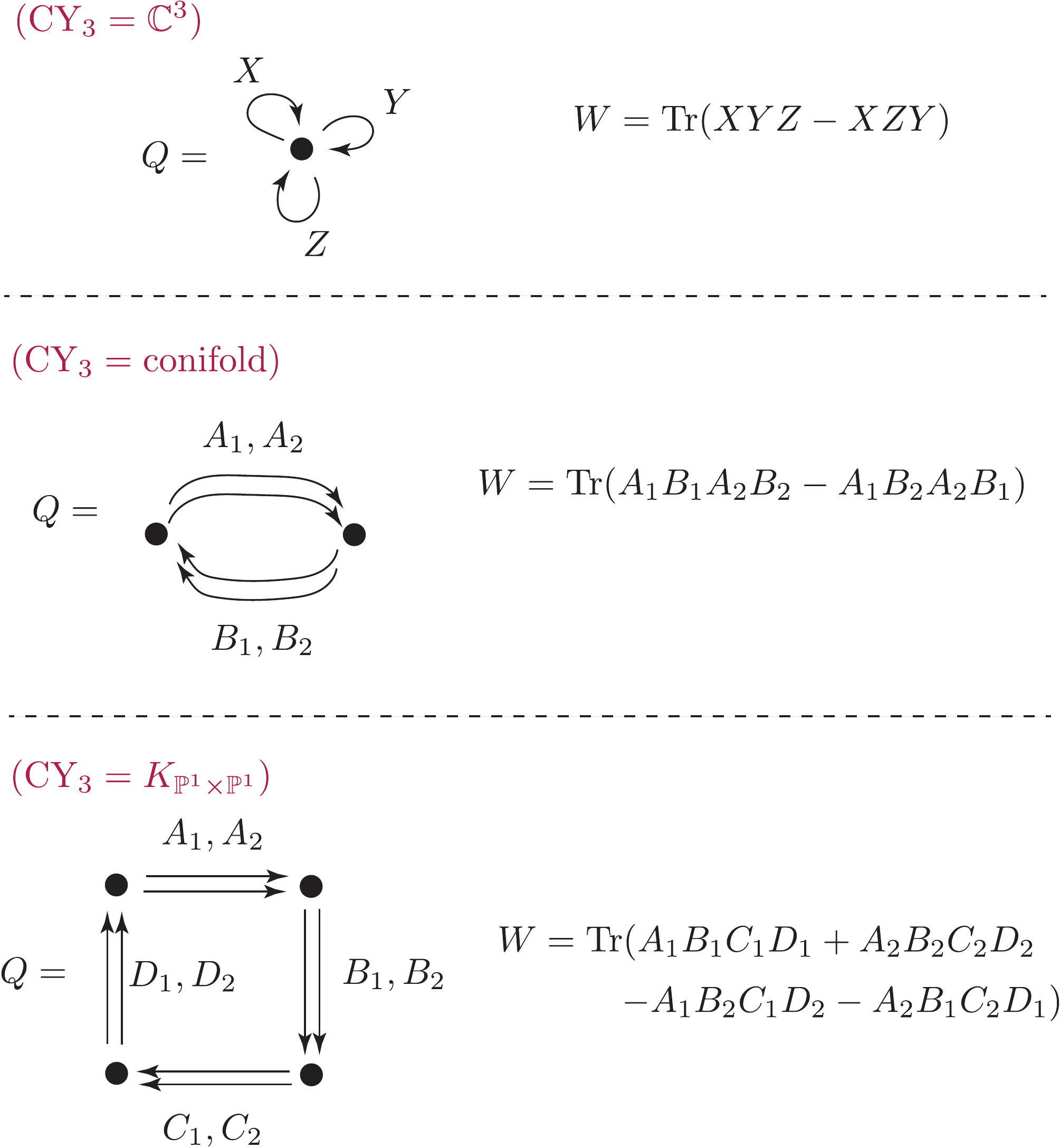}
\caption{Examples of quivers $Q$ and superpotentials $W$ describing toric Calabi-Yau three-folds,
$\mathbb{C}^3$, resolved conifold and the canonical bundle over $\mathbb{P}^1\times \mathbb{P}^1$.}
\label{fig.QW}
\end{figure}

While the definition of the quiver Yangian in itself works for a general choice of 
$Q$ and $W$, in the following we are interested in the cases
where $(Q, W)$ originates from a toric Calabi-Yau three-fold:
the $\mathcal{N}=4$ supersymmetric quiver quantum mechanics
defined from $(Q, W)$ reproduces the toric Calabi-Yau three-fold as the vacuum moduli space.
While there have been a lot of discussions on identifying the quiver/superpotential from the toric data\footnote{See e.g.\ reviews \cite{Kennaway:2007tq,Yamazaki:2008bt} and references therein.
}, it is enough for the purposes of this exposition to look at 
a few examples shown in Figure \ref{fig.QW}.

We assign equivariant parameters (to be called charges) $h_I$ to each arrow $I\in Q_1$
such that the total charge of (any monomial term of) the superpotential $W$ is zero.
(For example, for $\mathbb{C}^3$ in Figure \ref{fig.QW} we 
associate parameters $h_X, h_Y, h_Z$ to $X, Y, Z$ with one constraint $h_X+h_Y+ h_Z=0$.)
This gives a parameterization of the flavor symmetries of the theory.

\subsection{Generators and Relations} 
We are now ready to define the $K$-shifted quiver Yangian $\mathsf{Y}(Q, W)$ associated with $(Q, W)$, in terms of generators and relations:

\begin{itemize}

\item(Generators)

For each vertex $a$ of the quiver we have a triple  
$(e^{(a)}(z), f^{(a)}(z), \psi^{(a)}(z))$, whose mode expansion gives an infinite set of generators $(e_n^{(a)}, \psi_n^{(a)},  f_n^{(a)})$:
\begin{equation}\label{eq-mode-expansion-toric}
\begin{aligned}
e^{(a)}(z)\equiv\sum^{+\infty}_{n=0}\frac{e^{(a)}_n}{z^{n+1}} \;, \qquad
\psi^{(a)}(z)\equiv \sum^{+\infty}_{n=K}\frac{\psi^{(a)}_n}{z^{n+1}}\,, \qquad
 f^{(a)}(z)\equiv \sum^{+\infty}_{n=0}\frac{f^{(a)}_n}{z^{n+1}} \;,
\end{aligned}
\end{equation}
with $K$ being the ``shift''.

\item(Relations)

The generators satisfy the relations
\begin{align}\label{eq-OPE-toric}
\begin{aligned}
\psi^{(a)}(z)\, \psi^{(b)}(w)&=   \psi^{(b)}(w)\, \psi^{(a)}(z)  \;,\\
\psi^{(a)}(z)\, e^{(b)}(w)&\simeq  \varphi^{b\Rightarrow a}(z-w)\, e^{(b)}(w)\, \psi^{(a)}(z)  \;, \\
e^{(a)}(z)\, e^{(b)}(w)&\sim  (-1)^{|a||b|}  \varphi^{b\Rightarrow a}(z-w) \, e^{(b)}(w)\, e^{(a)}(z)  \;, \\
\psi^{(a)}(z)\, f^{(b)}(w)&\simeq   \varphi^{b\Rightarrow a}(z-w)^{-1} \, f^{(b)}(w)\,\psi^{(a)}(z) \;,\\
f^{(a)}(z)\, f^{(b)}(w)&\sim  (-1)^{|a||b|} \varphi^{b\Rightarrow a}(z-w)^{-1}\,  f^{(b)}(w)\, f^{(a)}(z)   \;,\\
\left[e^{(a)}(z),f^{(b)}(w) \right\} &\sim -  \delta^{a,b} \frac{\psi^{(a)}(z)-\psi^{(b)}(w)}{z-w}  \;.
\end{aligned}
\end{align}
Here the relations are understood to hold in the expansions in powers of $z$ and $w$, with ``$\simeq$" meaning equality up to $z^n w^{m\geq 0}$ terms,  and ``$\sim$" meaning equality up to $z^{n\geq 0} w^{m}$ and $z^{n} w^{m\geq 0}$ terms.
The bracket $[e^{(a)}(z),f^{(b)}(w)\}$ represents a supercommutator as determined by a $\mathbb{Z}_2$-grading
\begin{align}\label{eq.Z2_grading}
|a|=\begin{cases}
0 & (\textrm{if there exists } I\in Q_1 \textrm{ s.t. $I$ begins and ends at $a$}) \;,\\
1 & (\textrm{otherwise}) \;.
\end{cases}
\end{align}
The ``bond factor'' $\varphi^{a\Rightarrow b}(z)$ is defined to be\footnote{The signs in front are chosens such that 
$\varphi^{a\Rightarrow b} (u) \, \varphi^{b\Rightarrow a} (-u)=1$, 
which is needed for the consistency of the relations; see Ref.\ \onlinecite{Galakhov:2021vbo} for details. Alternatively we can disregard the signs by choosing an ordering between the vertices.}
\begin{align}\label{eq-charge-atob}
\varphi^{a\Rightarrow b} (u)\equiv \pm \frac{\prod_{I\in \{b\rightarrow a\}}(u+h_{I})}{\prod_{I\in \{a\rightarrow b\}}(u-h_{I})} \;,
\end{align}
\noindent where $\{a\rightarrow b\}$ denotes the set of edges from vertex $a$ to vertex $b$.

\end{itemize}

As shown in Ref.\ \onlinecite{Li:2020rij}, the quiver Yangians contain interesting algebras as special examples:

\begin{itemize}

\item For $\mathbb{C}^3$ we get $\mathsf{Y}(\widehat{\mathfrak{gl}}_1)$, 
the affine Yangian of $\mathfrak{gl}_1$.

\item For conifold, we get $\mathsf{Y}(\widehat{\mathfrak{gl}}_{1|1})$,
the affine Yangian of $\mathfrak{gl}_{1|1}$.

\item More generally, for a toric Calabi-Yau three-fold $xy=z^n w^m$ (often called a generalized conifold)
we get an affine Yangian of $\mathfrak{gl}_{m|n}$.
\footnote{See Ref.\ \onlinecite{Bezerra:2019dmp} for recent discussion on the trigonometric/quantum-toroidal counterparts.} Note that for Lie superalgebras there are ambiguities in
 the choice of the Cartan matrix, but the affine Yangian is independent of such a choice up to an isomorphism.\cite{2019arXiv191208729B}

\item For $(Q, W)$ corresponding to a general toric Calabi-Yau manifold
there is no associated Lie algebra $\mathfrak{g}$, 
and the algebra seems to be new.
In general, we have multiple quiver/superpotential pairs for the same toric Calabi-Yau geometry,
and it was conjectured in Ref.\ \onlinecite{Li:2020rij} that their quiver Yangians are all isomorphic.

\end{itemize}

\section{Representation from Crystal Melting}

\subsection{Crystal Melting}

\begin{figure}[htbp]
\includegraphics[scale=0.25]{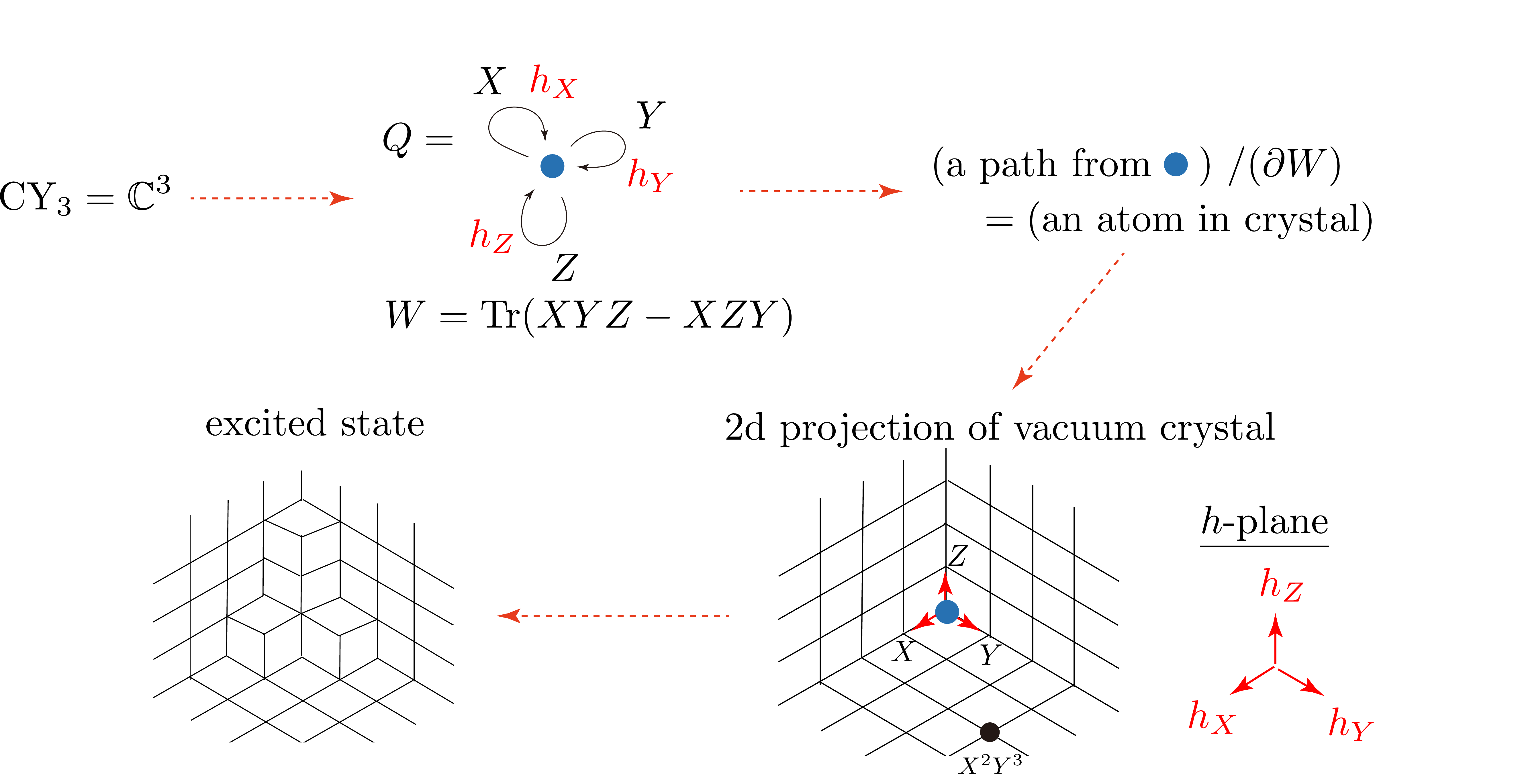}
\caption{The crystal melting model for $\mathbb{C}^3$.}
\label{fig.crystal}
\end{figure}

The concept of crystal melting is best explained in the simplest example of $\mathbb{C}^3$,\cite{Okounkov:2003sp,Iqbal:2003ds} whose quiver and superpotential are shown in Figure \ref{fig.crystal}.

We can describe the moduli space by associating three variables $X, Y, Z$ to the three edges of the quiver.
The superpotential constraints give $\partial W/ \partial X = YZ-ZY =0$, as well as its cyclic permutations.
This means that we have a polynomial ring $\mathbb{C}[X, Y, Z]$, which reproduces the coordinate ring of $\mathbb{C}^3$.
Now, to construct the crystal we consider all paths starting at the vertex modulo the superpotential constraints,
so that we obtain $X^i Y^j Z^k$ with $i,j,k \ge 0$. Let us call such an equivalence class of paths to be an atom.
By placing the atoms at the locations $(i,j,k)$
(namely, according to their charges\footnote{We obtain a two-dimensional projection of the crystal
when we consider flavor symmetries parametrized by $h_X, h_Y, h_Z$ with $h_X+h_Y+h_Z=0$. To obtain a three-dimensional crystal we also need to consider an R-symmetry, which amounts to lifting the condition $h_X+h_Y+h_Z=0$, so that we have three parameters $h_X, h_Y, h_Z$.
The atom located at $(i,j,k)$ has $h_X$ charge $i h_X$, $h_Y$ charge $j h_Y$ and $h_Z$ charge $k h_Z$.}), we obtain a three-dimensional arrangement of atoms, which we call the vacuum crystal. The excited states of the crystal are obtained by removing atoms from the corner of the crystal.

Now the observation of Ref.\ \onlinecite{Ooguri:2009ijd} (see also Ref.\ \onlinecite{MR2592501}) is that this story generalizes to an 
arbitrary toric Calabi-Yau manifold. We refer to Refs.\ \onlinecite{Ooguri:2009ijd,Yamazaki:2010fz}
for detailed explanation; we here instead show an example of a crystal,
since a picture is worth a thousand words (Figure \ref{fig.crystalSPP}):

\begin{figure}[htbp]
\includegraphics[scale=0.25]{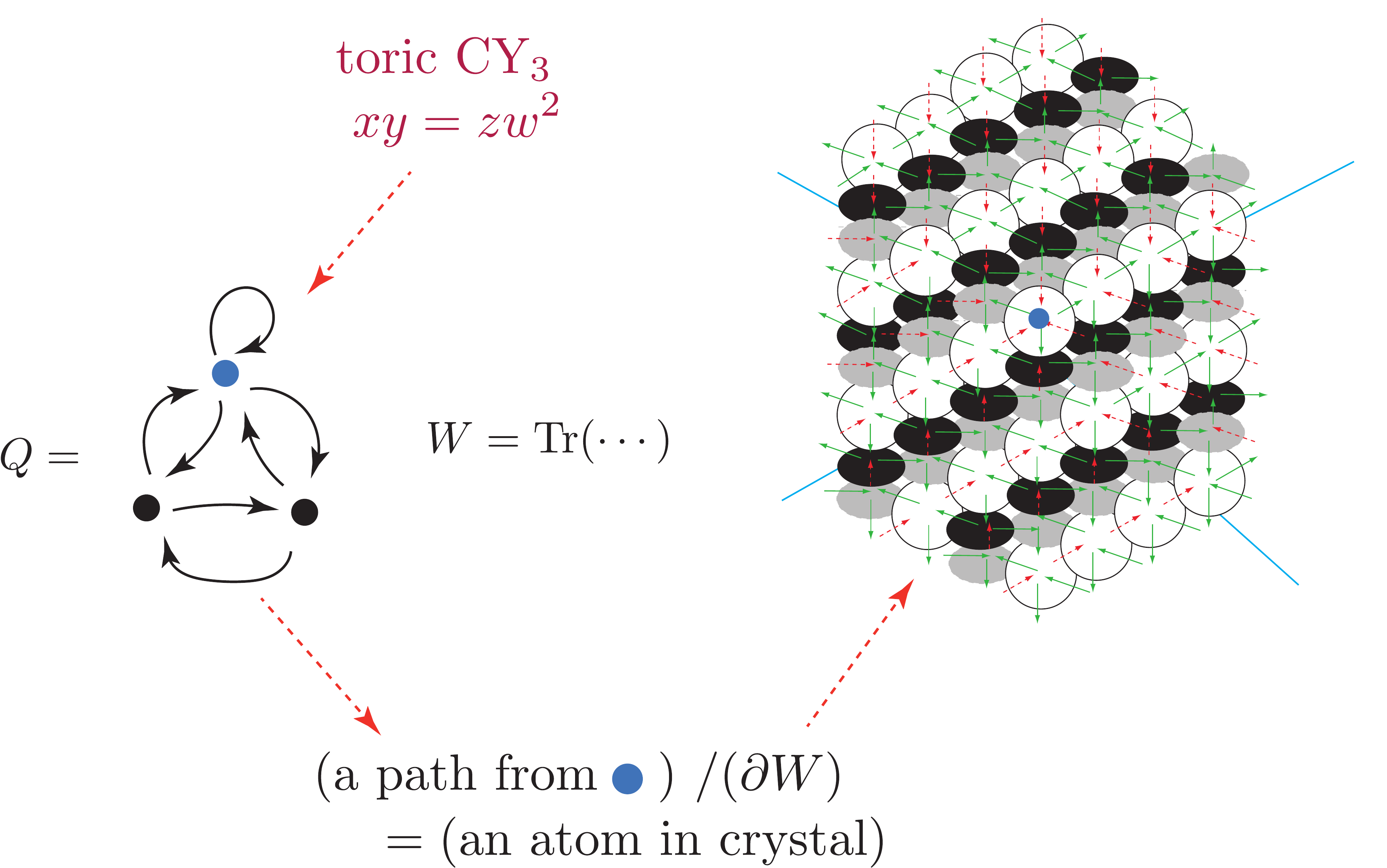}
\caption{An example of the vacuum crystal associated with the Calabi-Yau geometry $xy=zw^2$. The origin/top of the crystal is shown by the blue dot in the center. (Figures reproduced and modified from Ref.\ \onlinecite{Ooguri:2009ijd}.)}
\label{fig.crystalSPP}
\end{figure}

The crystal shown in Figure \ref{fig.crystalSPP} shows the ``vacuum'' of the model.
The excited states are given by ``molten crystals'', which are obtained by 
removing a finite set of atoms $\Lambda$ from the top of the crystal. 
This gives a generalization of the $\mathbb{C}^3$-crystal melting,
where the connectivities of the atoms are dictated by the edges of the quiver,
and the atoms are colored according to the vertices of the quiver 
(the end-point of the path corresponding to the atom).

\begin{figure}[htbp]
\centering\includegraphics[scale=0.13]{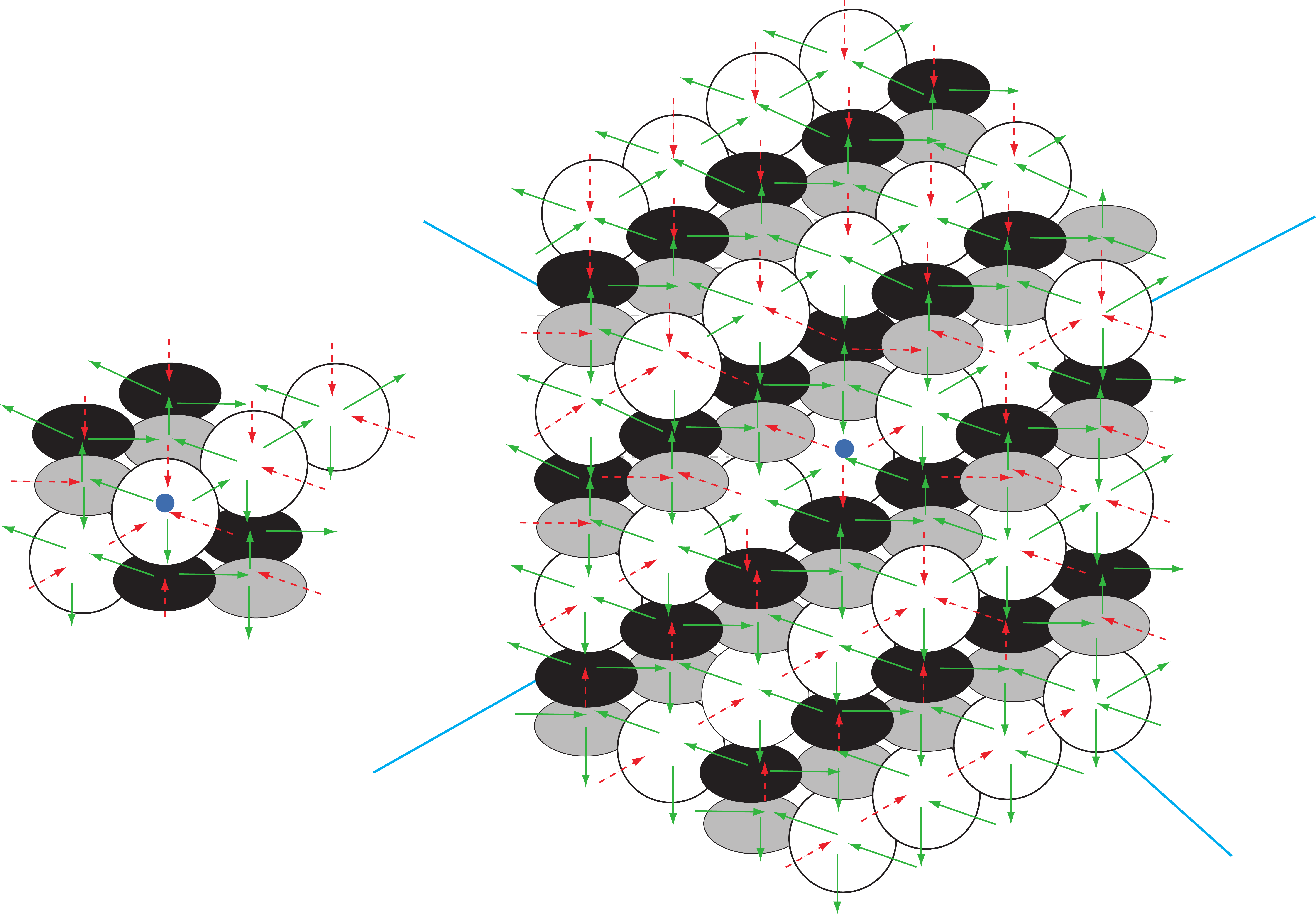}
\caption{An example of a configuration of the molten crystal (right) and the complement, the crystal configuration (left), for the crystal of 
Figure \ref{fig.crystalSPP}. (Figures reproduced from Ref.\ \onlinecite{Ooguri:2009ijd}.)}
\label{SPPmolten_rev}
\end{figure}

\subsection{Representations}

We can now describe a representation of the quiver Yangian.
The representation space is spanned by 
the set of the complements of molten crystal configurations: $|\Lambda\rangle$.
The generators of the quiver Yangian act on the states by the following formulas:\footnote{These formulas are inspired by their counterparts for $\mathbb{C}^3$, see Refs.\ \onlinecite{feigin2012, Tsymbaliuk:2014fvq, Prochazka:2015deb}.
Some more examples of toric Calabi-Yau manifolds also discussed e.g.\ in  Refs.\ \onlinecite{Feigin1204,Feigin:2013fga,Gaberdiel:2017dbk,Rapcak:2018nsl,Rapcak:2020ueh}.}
\begin{equation}\label{eq-ansatz-action-toric}
\begin{aligned}
\psi^{(a)}(z)|\Lambda\rangle&= \Psi_{\Lambda}^{(a)}(z)|\Lambda\rangle \;,\\
e^{(a)}(z)|\Lambda\rangle &=\sum_{\sqbox{$a$} \,\in \,\textrm{Add}(\Lambda)} 
 \frac{E^{(a)}(\Lambda\rightarrow \Lambda+\sqbox{$a$})}{z-h(\sqbox{$a$})}|\Lambda+\sqbox{$a$}\rangle \;,\\
f^{(a)}(z)|\Lambda\rangle &=\sum_{\sqbox{$a$}\, \in\, \textrm{Rem}(\Lambda)}
\frac{F^{(a)}(\Lambda\rightarrow \Lambda-\sqbox{$a$})}{z-h(\sqbox{$a$})}|\Lambda-\sqbox{$a$}\rangle \;.\\
\end{aligned}
\end{equation}

While $\psi^{(a)}(z)$ acts diagonally on the crystal basis,
$e^{(a)}(z)$ ($f^{(a)}(z)$) adds (removes) an atom
from the crystal: here $\sqbox{$a$} \in\textrm{Add}(\Lambda)$ ($\sqbox{$a$}\in\textrm{Rem}(\Lambda)$)) means that we consider an atom of color $a$ which can be added to (removed from) the crystal $\Lambda$.
Note that for each atom $\sqbox{$a$}$ which can be added to (removed from) the crystal there is a corresponding pole at $z=h(\sqbox{$a$})$ on the right-hand side.

The expressions $\Psi^{(a)}_{\Lambda}(u)$, 
$E^{(a)}(\Lambda\rightarrow \Lambda+\sqbox{$a$})$ and 
$F^{(a)}(\Lambda\rightarrow \Lambda-\sqbox{$a$})$ are given by
\begin{align}\label{eq-Psi-ansatz-toric}
&\Psi^{(a)}_{\Lambda}(u)=\psi^{(a)}_{0}(z)\prod_{b\in Q_0} \prod_{\sqbox{$b$}\in \Lambda} \varphi^{b\Rightarrow a}(u-h(\sqbox{$b$}))  \;, \\
&E^{(a)}(\Lambda\rightarrow \Lambda+\sqbox{$a$})= \pm \sqrt{\pm\textrm{Res}_{u=h(\sqbox{$a$})}\Psi^{(a)}_{\Lambda}(u)}\;, \\
&F^{(a)}(\Lambda\rightarrow \Lambda-\sqbox{$a$})=\pm \sqrt{\pm\textrm{Res}_{u=h(\sqbox{$a$})}\Psi^{(a)}_{\Lambda}(u)}\;,
\end{align}
where we here do not discuss subtle sign choices, for which readers are referred to Ref.\ \onlinecite{Li:2020rij}.
($E^{(a)}$ and $F^{(a)}$ are essentially the same expressions except for sign differences.)
Note that $\varphi^{a\Rightarrow b} (u)$ was defined previously in \eqref{eq-charge-atob}.

\begin{figure}[htbp]
\includegraphics[scale=0.37]{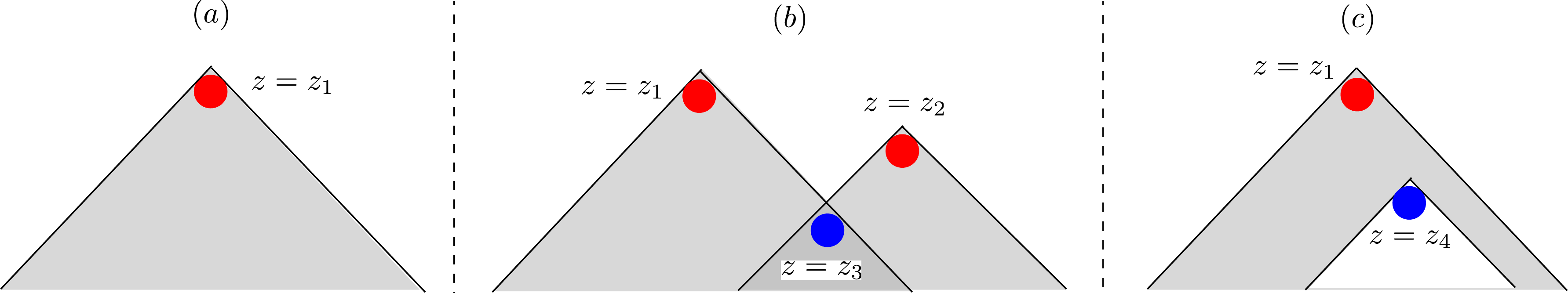}
\caption{The vacuum charge function $\psi_0^{(a)}$ determines the shape of the crystal.}
\label{fig.psi0}
\end{figure}

We still need to describe the function $\psi_0^{(a)}(z)$, which determines the action of the $\psi^{(a)}(z)$ generator on the vacuum crystal:
\begin{equation}
\begin{aligned}
\psi^{(a)}(z)|\varnothing \rangle&= \psi_{0}^{(a)}(z)|\varnothing \rangle \;.\\
\end{aligned}
\end{equation}
The simplest choice of $\psi_0^{(a)}(z)$ is to allow for a single-order pole for one of the vertices $a$,
so that we have $\psi_0^{(a)}(z) = 1/(z-z_1)$.
This corresponds to starting the growth of the crystal from the vertex $a$, at the location $z= z_1$ of the ``starter''.
We can consider a more complicated choice, for example $\psi_0^{(a)}(z)= (z-z_3)/(z-z_1)(z-z_2)$.
In this case, the growth of the crystal starts at two separate locations,
at $z= z_1$ and $z= z_2$; the two crystals then begin to overlap at $z=z_3$,
and the zero of $\psi_0^{(a)}(z)$ ensures that the two copies of the crystal tails should be identified beyond $z=z_3$ (the location of the ``pauser''),
so that we have a single crystal of the combined shape shown in Figure \ref{fig.psi0}.
We can also delete portions of the crystal as in Figure \ref{fig.psi0} (c), by 
considering the $\psi_0^{(a)}(z) = (z-z_4)/(z-z_1)$ where $z=z_4$ (the ``stopper'') is located inside the crystal.
In general, we can choose $\psi_0^{(a)}(z)$ to be a rational function---the poles of $\psi_0^{(a)}$ specify the 
``starters'' of the crystals, while its zeros specify the ``pauser/stoppers'' (which can be regarded as starters for ``negative crystals"); we obtain a representation of the $K$-shifted quiver Yangian, where $K$ is determined by a net degree of $\psi_0^{(a)}$ (degree of the numerator minus that of the denominator when we express $\psi_0^{(a)}$ as a ratio of two polynomials.).\footnote{In general we can play such a game for $\psi_0^{(a)}$ for each quiver vertex $a$.}

\begin{figure}[htbp]
\includegraphics[scale=0.23]{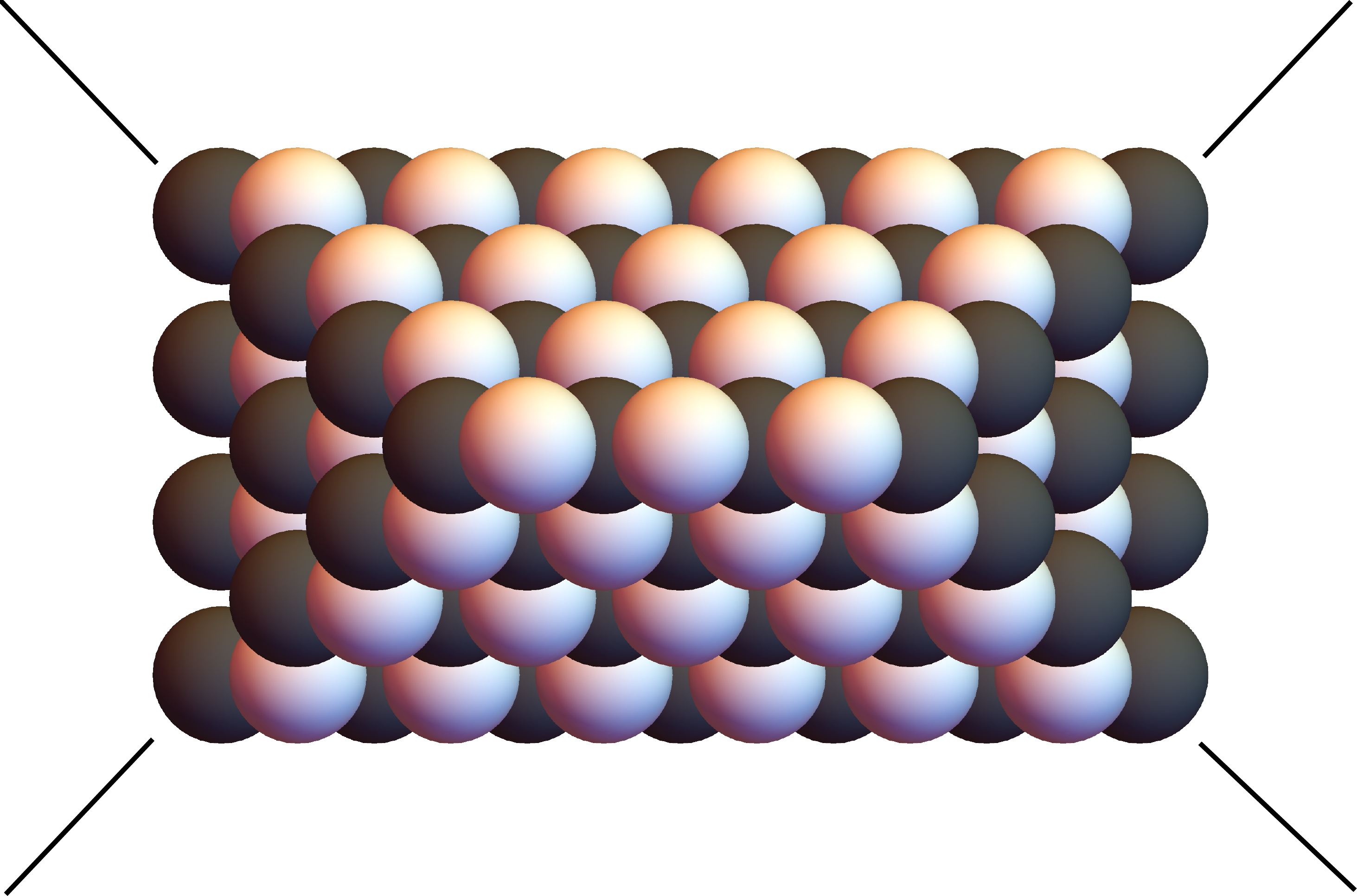}
\qquad
\includegraphics[scale=0.15]{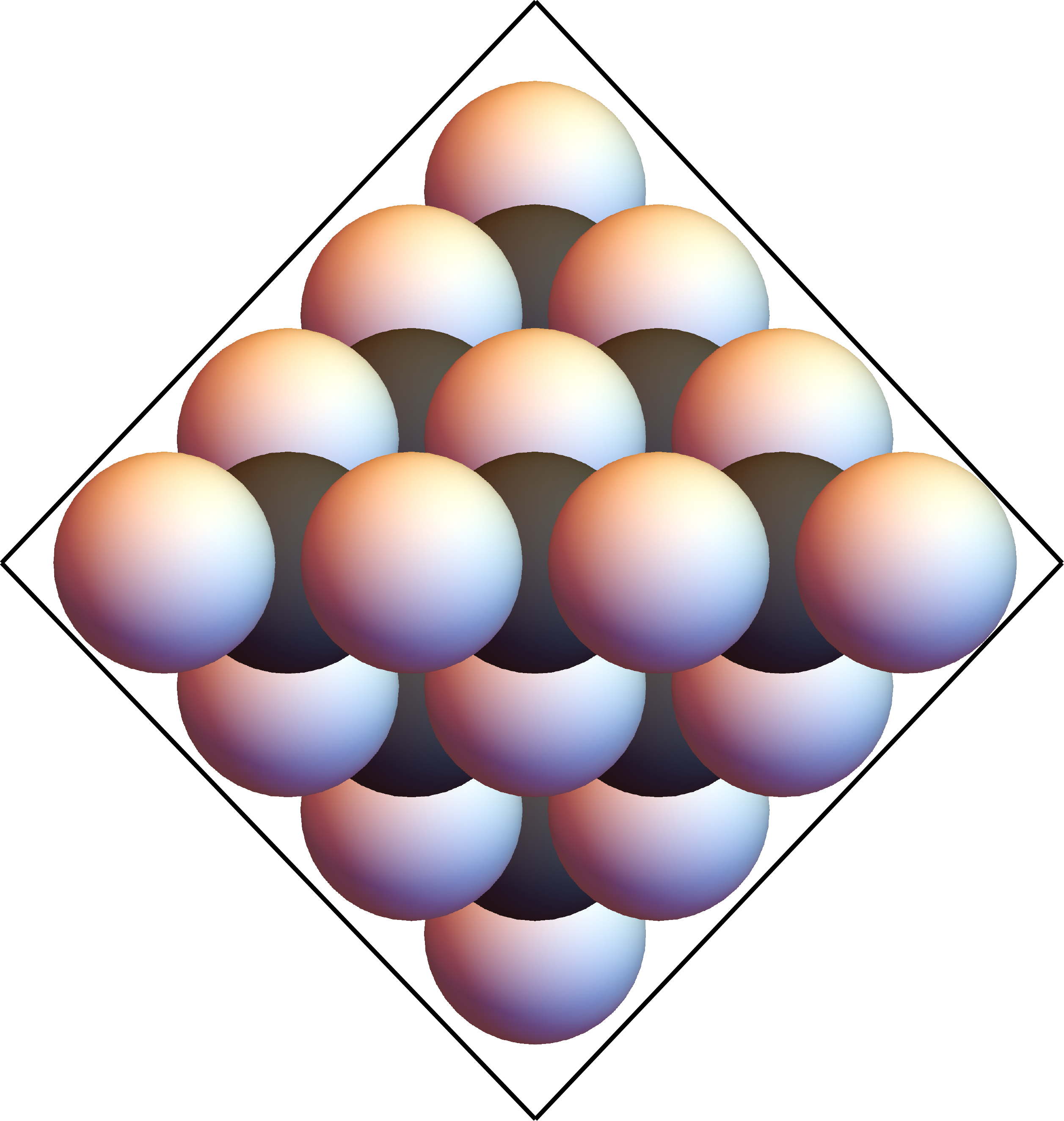}
\caption{Examples of crystals representing BPS state countings of the conifold, as discussed in 
Refs.\ \onlinecite{MR2836398,Jafferis:2008uf,Chuang:2008aw}. (Figures reproduced and modified from Ref.\ \onlinecite{Galakhov:2021xum}.)}
\label{fig.wall_crossing}
\end{figure}

Generalizing this argument, we can obtain
a rather general class of representations using 
a combination of starter/pauser/stoppers, and
a general shape of the crystal can be used for a representation.
This includes, for example, open/closed BPS state countings and 
their wall crossings.\cite{Okounkov:2003sp,MR2836398,Jafferis:2008uf,Chuang:2008aw,Nagao:2009ky,Nagao:2009rq,Yamazaki:2010fz,Sulkowski:2010eg} 
In Figure \ref{fig.wall_crossing}
we show examples of crystals relevant for wall crossings in the resolved conifold.

\begin{figure}[htbp]
\resizebox{.7\linewidth}{!}{
	\begin{tikzpicture}[scale=0.5]
	\node at (13,0) {\includegraphics[scale=0.2]{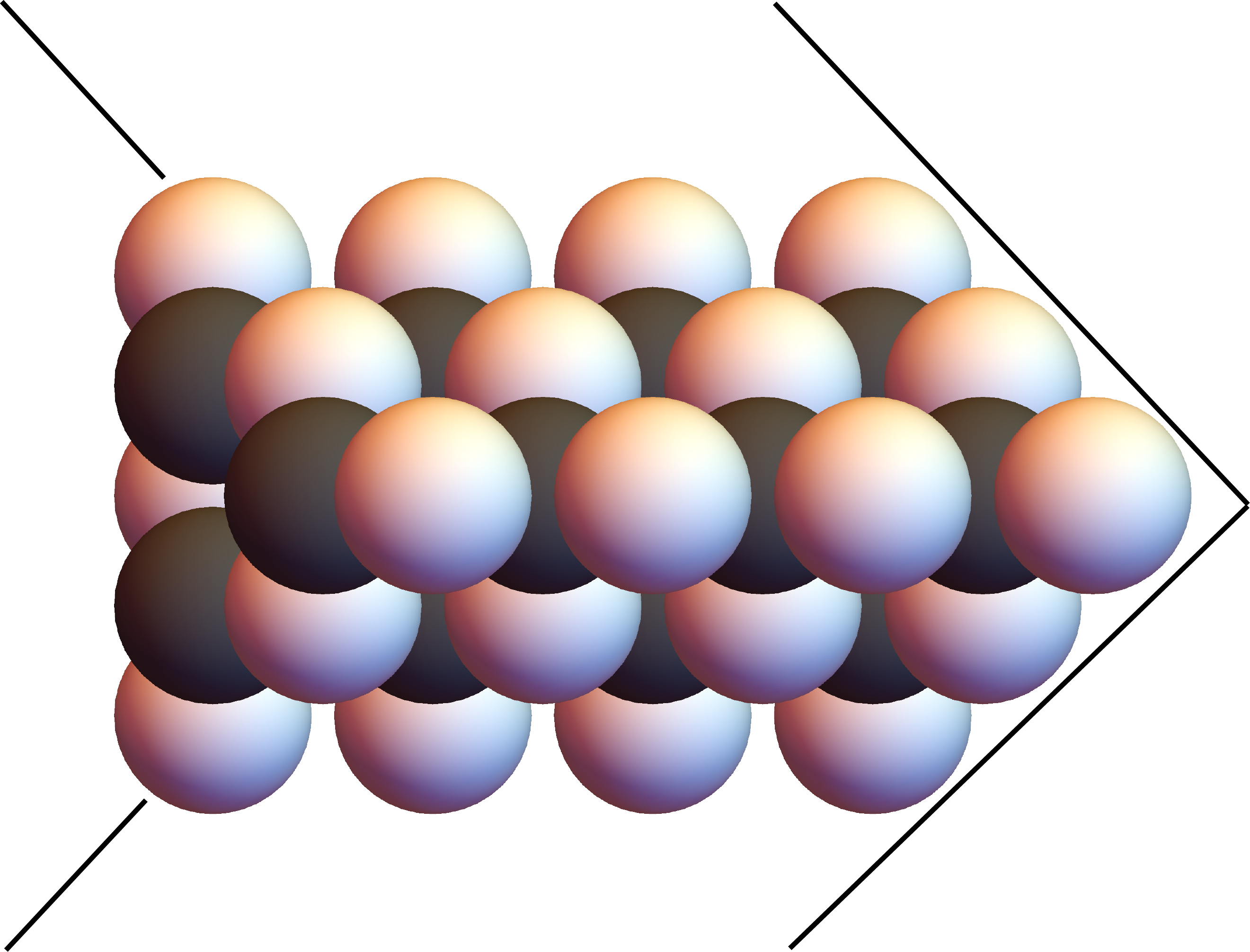}};
	
		\foreach \a in {0,...,4}
		\foreach \b in {0,...,2}
		\foreach \c in {3,...,3}
		{
			\draw[thick, fill=gray] (-0.866025*\a+0.866025*\b,-0.5*\a-0.5*\b+\c) -- (-0.866025*\a+0.866025*\b,-0.5*\a-0.5*\b+\c - 1) --(-0.866025*\a+0.866025*\b -0.866025,-0.5*\a-0.5*\b+\c -0.5) -- (-0.866025*\a+0.866025*\b -0.866025,-0.5*\a-0.5*\b+\c +0.5) -- cycle
			(-0.866025*\a+0.866025*\b,-0.5*\a-0.5*\b+\c) -- (-0.866025*\a+0.866025*\b,-0.5*\a-0.5*\b+\c - 1) --(-0.866025*\a+0.866025*\b +0.866025,-0.5*\a-0.5*\b+\c -0.5) -- (-0.866025*\a+0.866025*\b +0.866025,-0.5*\a-0.5*\b+\c +0.5) -- cycle
			(-0.866025*\a+0.866025*\b,-0.5*\a-0.5*\b+\c) -- (-0.866025*\a+0.866025*\b -0.866025,-0.5*\a-0.5*\b+\c +0.5) -- (-0.866025*\a+0.866025*\b,-0.5*\a-0.5*\b+\c+1) -- (-0.866025*\a+0.866025*\b +0.866025,-0.5*\a-0.5*\b+\c +0.5) -- cycle;
		}
	\foreach \a in {0,...,4}
	\foreach \b in {2,...,2}
	\foreach \c in {0,...,3}
	{
		\draw[thick, fill=gray] (-0.866025*\a+0.866025*\b,-0.5*\a-0.5*\b+\c) -- (-0.866025*\a+0.866025*\b,-0.5*\a-0.5*\b+\c - 1) --(-0.866025*\a+0.866025*\b -0.866025,-0.5*\a-0.5*\b+\c -0.5) -- (-0.866025*\a+0.866025*\b -0.866025,-0.5*\a-0.5*\b+\c +0.5) -- cycle
		(-0.866025*\a+0.866025*\b,-0.5*\a-0.5*\b+\c) -- (-0.866025*\a+0.866025*\b,-0.5*\a-0.5*\b+\c - 1) --(-0.866025*\a+0.866025*\b +0.866025,-0.5*\a-0.5*\b+\c -0.5) -- (-0.866025*\a+0.866025*\b +0.866025,-0.5*\a-0.5*\b+\c +0.5) -- cycle
		(-0.866025*\a+0.866025*\b,-0.5*\a-0.5*\b+\c) -- (-0.866025*\a+0.866025*\b -0.866025,-0.5*\a-0.5*\b+\c +0.5) -- (-0.866025*\a+0.866025*\b,-0.5*\a-0.5*\b+\c+1) -- (-0.866025*\a+0.866025*\b +0.866025,-0.5*\a-0.5*\b+\c +0.5) -- cycle;
	}
\foreach \a in {4,...,4}
\foreach \b in {0,...,2}
\foreach \c in {0,...,3}
{
	\draw[thick, fill=gray] (-0.866025*\a+0.866025*\b,-0.5*\a-0.5*\b+\c) -- (-0.866025*\a+0.866025*\b,-0.5*\a-0.5*\b+\c - 1) --(-0.866025*\a+0.866025*\b -0.866025,-0.5*\a-0.5*\b+\c -0.5) -- (-0.866025*\a+0.866025*\b -0.866025,-0.5*\a-0.5*\b+\c +0.5) -- cycle
	(-0.866025*\a+0.866025*\b,-0.5*\a-0.5*\b+\c) -- (-0.866025*\a+0.866025*\b,-0.5*\a-0.5*\b+\c - 1) --(-0.866025*\a+0.866025*\b +0.866025,-0.5*\a-0.5*\b+\c -0.5) -- (-0.866025*\a+0.866025*\b +0.866025,-0.5*\a-0.5*\b+\c +0.5) -- cycle
	(-0.866025*\a+0.866025*\b,-0.5*\a-0.5*\b+\c) -- (-0.866025*\a+0.866025*\b -0.866025,-0.5*\a-0.5*\b+\c +0.5) -- (-0.866025*\a+0.866025*\b,-0.5*\a-0.5*\b+\c+1) -- (-0.866025*\a+0.866025*\b +0.866025,-0.5*\a-0.5*\b+\c +0.5) -- cycle;
}
\foreach \a in {5,...,5}
\foreach \b in {0,...,0}
\foreach \c in {0,...,0}
{
	\draw[thick, fill=red] (-0.866025*\a+0.866025*\b,-0.5*\a-0.5*\b+\c) -- (-0.866025*\a+0.866025*\b,-0.5*\a-0.5*\b+\c - 1) --(-0.866025*\a+0.866025*\b -0.866025,-0.5*\a-0.5*\b+\c -0.5) -- (-0.866025*\a+0.866025*\b -0.866025,-0.5*\a-0.5*\b+\c +0.5) -- cycle
	(-0.866025*\a+0.866025*\b,-0.5*\a-0.5*\b+\c) -- (-0.866025*\a+0.866025*\b,-0.5*\a-0.5*\b+\c - 1) --(-0.866025*\a+0.866025*\b +0.866025,-0.5*\a-0.5*\b+\c -0.5) -- (-0.866025*\a+0.866025*\b +0.866025,-0.5*\a-0.5*\b+\c +0.5) -- cycle
	(-0.866025*\a+0.866025*\b,-0.5*\a-0.5*\b+\c) -- (-0.866025*\a+0.866025*\b -0.866025,-0.5*\a-0.5*\b+\c +0.5) -- (-0.866025*\a+0.866025*\b,-0.5*\a-0.5*\b+\c+1) -- (-0.866025*\a+0.866025*\b +0.866025,-0.5*\a-0.5*\b+\c +0.5) -- cycle;
}
\foreach \a in {0,...,0}
\foreach \b in {3,...,3}
\foreach \c in {0,...,0}
{
	\draw[thick, fill=red] (-0.866025*\a+0.866025*\b,-0.5*\a-0.5*\b+\c) -- (-0.866025*\a+0.866025*\b,-0.5*\a-0.5*\b+\c - 1) --(-0.866025*\a+0.866025*\b -0.866025,-0.5*\a-0.5*\b+\c -0.5) -- (-0.866025*\a+0.866025*\b -0.866025,-0.5*\a-0.5*\b+\c +0.5) -- cycle
	(-0.866025*\a+0.866025*\b,-0.5*\a-0.5*\b+\c) -- (-0.866025*\a+0.866025*\b,-0.5*\a-0.5*\b+\c - 1) --(-0.866025*\a+0.866025*\b +0.866025,-0.5*\a-0.5*\b+\c -0.5) -- (-0.866025*\a+0.866025*\b +0.866025,-0.5*\a-0.5*\b+\c +0.5) -- cycle
	(-0.866025*\a+0.866025*\b,-0.5*\a-0.5*\b+\c) -- (-0.866025*\a+0.866025*\b -0.866025,-0.5*\a-0.5*\b+\c +0.5) -- (-0.866025*\a+0.866025*\b,-0.5*\a-0.5*\b+\c+1) -- (-0.866025*\a+0.866025*\b +0.866025,-0.5*\a-0.5*\b+\c +0.5) -- cycle;
}
\foreach \a in {0,...,0}
\foreach \b in {0,...,0}
\foreach \c in {4,...,4}
{
	\draw[thick, fill=red] (-0.866025*\a+0.866025*\b,-0.5*\a-0.5*\b+\c) -- (-0.866025*\a+0.866025*\b,-0.5*\a-0.5*\b+\c - 1) --(-0.866025*\a+0.866025*\b -0.866025,-0.5*\a-0.5*\b+\c -0.5) -- (-0.866025*\a+0.866025*\b -0.866025,-0.5*\a-0.5*\b+\c +0.5) -- cycle
	(-0.866025*\a+0.866025*\b,-0.5*\a-0.5*\b+\c) -- (-0.866025*\a+0.866025*\b,-0.5*\a-0.5*\b+\c - 1) --(-0.866025*\a+0.866025*\b +0.866025,-0.5*\a-0.5*\b+\c -0.5) -- (-0.866025*\a+0.866025*\b +0.866025,-0.5*\a-0.5*\b+\c +0.5) -- cycle
	(-0.866025*\a+0.866025*\b,-0.5*\a-0.5*\b+\c) -- (-0.866025*\a+0.866025*\b -0.866025,-0.5*\a-0.5*\b+\c +0.5) -- (-0.866025*\a+0.866025*\b,-0.5*\a-0.5*\b+\c+1) -- (-0.866025*\a+0.866025*\b +0.866025,-0.5*\a-0.5*\b+\c +0.5) -- cycle;
}
\node[left] at (-4.33013, -0.5) {$k$ atoms};
\node[above left] at (-2.16506, 2.75) {$m$ atoms};
\node[above right] at (1.29904, 3.25) {$n$ atoms};
	\end{tikzpicture}
}
\caption{Examples of the crystals representations with no known BPS state counting interpretations.\cite{Galakhov:2021xum} (Figures reproduced and modified from Ref.\ \onlinecite{Galakhov:2021xum}.)}
\label{fig.no_known_geometry}
\end{figure}
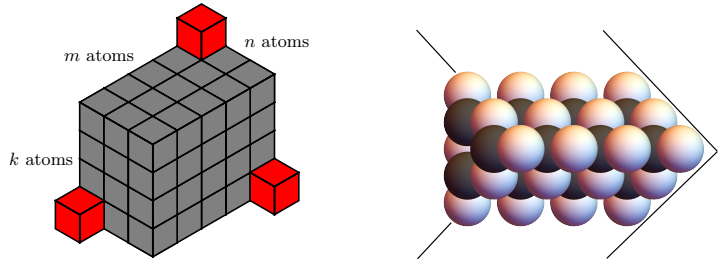

One should keep in mind, however, that almost all of the 
crystals representations have no known counterparts in the 
geometric discussions of Donaldson-Thomas invariants, 
see Figure \ref{fig.no_known_geometry} for examples of such crystals.

Finally, let us mention that the representations discussed above can become reducible for non-generic values of the equivariant parameters. In this case, we can truncate the algebra (namely, consider the quotient of the algebra by the ideal annihilating the null state), so that the representation is irreducible again.
This method gives rise to a zoo of interesting algebras. For $\mathbb{C}^3$ this gives  
vertex algebras discussed in Ref.\ \onlinecite{Gaiotto:2017euk}.

\subsection{Derivations from Supersymmetric Quiver Quantum Mechanics}

We have introduced quiver Yangians and their representations 
by a top-down approach. One can, however, {\it derive} 
the algebra by assuming the crystal representation \eqref{eq-ansatz-action-toric}
as an ansatz.\cite{Li:2020rij, Galakhov:2021xum} 
Alternatively, we can derive the algebras and their representations by an equivariant localization of supersymmetric quiver quantum mechanics.\cite{Galakhov:2020vyb,Galakhov:2021xum} In this section we explain the latter approach.

Let us here quickly summarize the salient features of the derivation (see Refs.\ \onlinecite{Galakhov:2020vyb, Galakhov:2021xum} for details). One starts with an $\mathcal{N}=4$ supersymmetric quiver quantum mechanics
defined from the quiver $Q$ and the superpotential $W$. 
By applying an equivariant localization to the Higgs branch of the moduli space,
the BPS state counting problem of the quiver quantum mechanics reduces to the counting of the fixed points,
which are captured by the configurations $\Lambda$ of crystal melting.
We can moreover identify the associated effective wavefunction $\Psi_{\Lambda}$ as the Euler class of the moduli space $\mathcal{M}_{\Lambda}$. Now, the action of the raising operator $e$ is to bring in an extra atom from infinity, 
causing the change of the wavefunction $\Psi_{\Lambda}$ into $\Psi_{\Lambda + \Box}$. 
This operation---the ``Hecke modification'' of the associated sheaf---is described by a Fourier-Mukai transformation over $\mathcal{M}_{\Lambda} \times \mathcal{M}_{\Lambda+\Box}$, whose kernel is given by the incidence relation. What is important in practice is that one can then explicitly compute the matrix elements of the generators, and verify the relations satisfied by the generators.\footnote{In general, in addition to \eqref{eq-OPE-toric} there are extra relations---Serre relations---satisfied by the generators. When the Serre relations are included we have the reduced quiver Yangian.}
This line of logic gives a first-principle derivation of the results and makes the connection with geometry more manifest. In practice, the match with quiver Yangians involves miraculous-looking cancellations as in Appendix of Ref.\ \onlinecite{Galakhov:2020vyb}.

\section{Summary}

Integrable models have a rather long history with so many sophisticated results in the literature.
However, the recent developments described in this paper demonstrate (rather convincingly in my opinion) 
that there are still new integrable structures yet to be discovered, even at the level of algebras generalizing Yangians.
I am optimistic that such a research direction will continue to prosper in the future, and that dialogues between 
physics and mathematics will be the crucial driving force in such an endeavor. 

\begin{acknowledgments}
The author would like to thank  Dimitry Galakhov and Wei Li for collaborations in the 
works\cite{Li:2020rij, Galakhov:2020vyb, Galakhov:2021xum} presented in this paper. He would also like to thank the organizers of the ICMP 2021 conference for the kind invitation.
The work of MY was supported in part
by WPI Research Center Initiative, MEXT, Japan and 
the JSPS Grant-in-Aid for Scientific Research (17KK0087, 19K03820, 19H00689,
20H05850, 20H05860).
\end{acknowledgments}

\bibliography{quiverYangian_ICMP}

\begin{thebibliography}{42}%
\makeatletter
\providecommand \@ifxundefined [1]{%
 \@ifx{#1\undefined}
}%
\providecommand \@ifnum [1]{%
 \ifnum #1\expandafter \@firstoftwo
 \else \expandafter \@secondoftwo
 \fi
}%
\providecommand \@ifx [1]{%
 \ifx #1\expandafter \@firstoftwo
 \else \expandafter \@secondoftwo
 \fi
}%
\providecommand \natexlab [1]{#1}%
\providecommand \enquote  [1]{``#1''}%
\providecommand \bibnamefont  [1]{#1}%
\providecommand \bibfnamefont [1]{#1}%
\providecommand \citenamefont [1]{#1}%
\providecommand \href@noop [0]{\@secondoftwo}%
\providecommand \href [0]{\begingroup \@sanitize@url \@href}%
\providecommand \@href[1]{\@@startlink{#1}\@@href}%
\providecommand \@@href[1]{\endgroup#1\@@endlink}%
\providecommand \@sanitize@url [0]{\catcode `\\12\catcode `\$12\catcode
  `\&12\catcode `\#12\catcode `\^12\catcode `\_12\catcode `\%12\relax}%
\providecommand \@@startlink[1]{}%
\providecommand \@@endlink[0]{}%
\providecommand \url  [0]{\begingroup\@sanitize@url \@url }%
\providecommand \@url [1]{\endgroup\@href {#1}{\urlprefix }}%
\providecommand \urlprefix  [0]{URL }%
\providecommand \Eprint [0]{\href }%
\providecommand \doibase [0]{http://dx.doi.org/}%
\providecommand \selectlanguage [0]{\@gobble}%
\providecommand \bibinfo  [0]{\@secondoftwo}%
\providecommand \bibfield  [0]{\@secondoftwo}%
\providecommand \translation [1]{[#1]}%
\providecommand \BibitemOpen [0]{}%
\providecommand \bibitemStop [0]{}%
\providecommand \bibitemNoStop [0]{.\EOS\space}%
\providecommand \EOS [0]{\spacefactor3000\relax}%
\providecommand \BibitemShut  [1]{\csname bibitem#1\endcsname}%
\let\auto@bib@innerbib\@empty
\bibitem [{Note1()}]{Note1}%
  \BibitemOpen
  \bibinfo {note} {See also Ref.\ \protect \rev@citealpnum {Rapcak:2021hdh} for
  another introduction on related topics.}\BibitemShut {Stop}%
\bibitem [{\citenamefont {Li}\ and\ \citenamefont
  {Yamazaki}(2020)}]{Li:2020rij}%
  \BibitemOpen
  \bibfield  {author} {\bibinfo {author} {\bibfnamefont {W.}~\bibnamefont
  {Li}}\ and\ \bibinfo {author} {\bibfnamefont {M.}~\bibnamefont {Yamazaki}},\
  }\bibfield  {title} {\enquote {\bibinfo {title} {{Quiver Yangian from Crystal
  Melting}},}\ }\href {\doibase 10.1007/JHEP11(2020)035} {\bibfield  {journal}
  {\bibinfo  {journal} {JHEP}\ }\textbf {\bibinfo {volume} {11}},\ \bibinfo
  {pages} {035} (\bibinfo {year} {2020})},\ \Eprint
  {http://arxiv.org/abs/2003.08909} {arXiv:2003.08909 [hep-th]} \BibitemShut
  {NoStop}%
\bibitem [{\citenamefont {Galakhov}\ and\ \citenamefont
  {Yamazaki}(2022)}]{Galakhov:2020vyb}%
  \BibitemOpen
  \bibfield  {author} {\bibinfo {author} {\bibfnamefont {D.}~\bibnamefont
  {Galakhov}}\ and\ \bibinfo {author} {\bibfnamefont {M.}~\bibnamefont
  {Yamazaki}},\ }\bibfield  {title} {\enquote {\bibinfo {title} {{Quiver
  Yangian and Supersymmetric Quantum Mechanics}},}\ }\href {\doibase
  10.1007/s00220-022-04490-y} {\bibfield  {journal} {\bibinfo  {journal}
  {Commun. Math. Phys.}\ }\textbf {\bibinfo {volume} {396}},\ \bibinfo {pages}
  {713--785} (\bibinfo {year} {2022})},\ \Eprint
  {http://arxiv.org/abs/2008.07006} {arXiv:2008.07006 [hep-th]} \BibitemShut
  {NoStop}%
\bibitem [{\citenamefont {Galakhov}, \citenamefont {Li},\ and\ \citenamefont
  {Yamazaki}(2021)}]{Galakhov:2021xum}%
  \BibitemOpen
  \bibfield  {author} {\bibinfo {author} {\bibfnamefont {D.}~\bibnamefont
  {Galakhov}}, \bibinfo {author} {\bibfnamefont {W.}~\bibnamefont {Li}}, \ and\
  \bibinfo {author} {\bibfnamefont {M.}~\bibnamefont {Yamazaki}},\ }\bibfield
  {title} {\enquote {\bibinfo {title} {{Shifted quiver Yangians and
  representations from BPS crystals}},}\ }\href {\doibase
  10.1007/JHEP08(2021)146} {\bibfield  {journal} {\bibinfo  {journal} {JHEP}\
  }\textbf {\bibinfo {volume} {08}},\ \bibinfo {pages} {146} (\bibinfo {year}
  {2021})},\ \Eprint {http://arxiv.org/abs/2106.01230} {arXiv:2106.01230
  [hep-th]} \BibitemShut {NoStop}%
\bibitem [{Note2()}]{Note2}%
  \BibitemOpen
  \bibinfo {note} {While we concentrate on the rational cases of quiver
  Yangians, the story generalizes to trigonometric\cite
  {Noshita:2021ldl,Galakhov:2021vbo,Noshita:2021dgj} and elliptic\cite
  {Galakhov:2021vbo} counterparts.}\BibitemShut {Stop}%
\bibitem [{\citenamefont {Drinfeld}(1985)}]{Drinfeld:1985rx}%
  \BibitemOpen
  \bibfield  {author} {\bibinfo {author} {\bibfnamefont {V.~G.}\ \bibnamefont
  {Drinfeld}},\ }\bibfield  {title} {\enquote {\bibinfo {title} {Hopf algebras
  and the quantum {Y}ang-{B}axter equation},}\ }\href@noop {} {\bibfield
  {journal} {\bibinfo  {journal} {Dokl. Akad. Nauk SSSR}\ }\textbf {\bibinfo
  {volume} {283}},\ \bibinfo {pages} {1060--1064} (\bibinfo {year}
  {1985})}\BibitemShut {NoStop}%
\bibitem [{\citenamefont {Drinfeld}(1986)}]{Drinfeld:1986}%
  \BibitemOpen
  \bibfield  {author} {\bibinfo {author} {\bibfnamefont {V.~G.}\ \bibnamefont
  {Drinfeld}},\ }\bibfield  {title} {\enquote {\bibinfo {title} {{Quantum
  groups}},}\ }\href@noop {} {\bibfield  {journal} {\bibinfo  {journal}
  {Proceedings of the International Congress of Mathematics}\ } (\bibinfo
  {year} {1986})}\BibitemShut {NoStop}%
\bibitem [{Note3()}]{Note3}%
  \BibitemOpen
  \bibinfo {note} {See e.g.\ reviews \cite {Kennaway:2007tq,Yamazaki:2008bt}
  and references therein.}\BibitemShut {Stop}%
\bibitem [{Note4()}]{Note4}%
  \BibitemOpen
  \bibinfo {note} {The signs in front are chosens such that $\varphi
  ^{a\Rightarrow b} (u) \protect \, \varphi ^{b\Rightarrow a} (-u)=1$, which is
  needed for the consistency of the relations; see Ref.\ \protect
  \rev@citealpnum {Galakhov:2021vbo} for details. Alternatively we can
  disregard the signs by choosing an ordering between the
  vertices.}\BibitemShut {Stop}%
\bibitem [{Note5()}]{Note5}%
  \BibitemOpen
  \bibinfo {note} {See Ref.\ \protect \rev@citealpnum {Bezerra:2019dmp} for
  recent discussion on the trigonometric/quantum-toroidal
  counterparts.}\BibitemShut {Stop}%
\bibitem [{\citenamefont {{Bezerra}}\ and\ \citenamefont
  {{Mukhin}}(2019)}]{2019arXiv191208729B}%
  \BibitemOpen
  \bibfield  {author} {\bibinfo {author} {\bibfnamefont {L.}~\bibnamefont
  {{Bezerra}}}\ and\ \bibinfo {author} {\bibfnamefont {E.}~\bibnamefont
  {{Mukhin}}},\ }\bibfield  {title} {\enquote {\bibinfo {title} {{Braid actions
  on quantum toroidal superalgebras}},}\ }\href@noop {} {\  (\bibinfo {year}
  {2019})},\ \Eprint {http://arxiv.org/abs/1912.08729} {arXiv:1912.08729
  [math.QA]} \BibitemShut {NoStop}%
\bibitem [{\citenamefont {Okounkov}, \citenamefont {Reshetikhin},\ and\
  \citenamefont {Vafa}(2006)}]{Okounkov:2003sp}%
  \BibitemOpen
  \bibfield  {author} {\bibinfo {author} {\bibfnamefont {A.}~\bibnamefont
  {Okounkov}}, \bibinfo {author} {\bibfnamefont {N.}~\bibnamefont
  {Reshetikhin}}, \ and\ \bibinfo {author} {\bibfnamefont {C.}~\bibnamefont
  {Vafa}},\ }\bibfield  {title} {\enquote {\bibinfo {title} {{Quantum
  Calabi-Yau and classical crystals}},}\ }\href {\doibase
  10.1007/0-8176-4467-9_16} {\bibfield  {journal} {\bibinfo  {journal} {Prog.
  Math.}\ }\textbf {\bibinfo {volume} {244}},\ \bibinfo {pages} {597} (\bibinfo
  {year} {2006})},\ \Eprint {http://arxiv.org/abs/hep-th/0309208}
  {arXiv:hep-th/0309208} \BibitemShut {NoStop}%
\bibitem [{\citenamefont {Iqbal}\ \emph {et~al.}(2008)\citenamefont {Iqbal},
  \citenamefont {Nekrasov}, \citenamefont {Okounkov},\ and\ \citenamefont
  {Vafa}}]{Iqbal:2003ds}%
  \BibitemOpen
  \bibfield  {author} {\bibinfo {author} {\bibfnamefont {A.}~\bibnamefont
  {Iqbal}}, \bibinfo {author} {\bibfnamefont {N.}~\bibnamefont {Nekrasov}},
  \bibinfo {author} {\bibfnamefont {A.}~\bibnamefont {Okounkov}}, \ and\
  \bibinfo {author} {\bibfnamefont {C.}~\bibnamefont {Vafa}},\ }\bibfield
  {title} {\enquote {\bibinfo {title} {{Quantum foam and topological
  strings}},}\ }\href {\doibase 10.1088/1126-6708/2008/04/011} {\bibfield
  {journal} {\bibinfo  {journal} {JHEP}\ }\textbf {\bibinfo {volume} {04}},\
  \bibinfo {pages} {011} (\bibinfo {year} {2008})},\ \Eprint
  {http://arxiv.org/abs/hep-th/0312022} {arXiv:hep-th/0312022} \BibitemShut
  {NoStop}%
\bibitem [{Note6()}]{Note6}%
  \BibitemOpen
  \bibinfo {note} {We obtain a two-dimensional projection of the crystal when
  we consider flavor symmetries parametrized by $h_X, h_Y, h_Z$ with
  $h_X+h_Y+h_Z=0$. To obtain a three-dimensional crystal we also need to
  consider an R-symmetry, which amounts to lifting the condition
  $h_X+h_Y+h_Z=0$, so that we have three parameters $h_X, h_Y, h_Z$. The atom
  located at $(i,j,k)$ has $h_X$ charge $i h_X$, $h_Y$ charge $j h_Y$ and $h_Z$
  charge $k h_Z$.}\BibitemShut {Stop}%
\bibitem [{\citenamefont {Ooguri}\ and\ \citenamefont
  {Yamazaki}(2009)}]{Ooguri:2009ijd}%
  \BibitemOpen
  \bibfield  {author} {\bibinfo {author} {\bibfnamefont {H.}~\bibnamefont
  {Ooguri}}\ and\ \bibinfo {author} {\bibfnamefont {M.}~\bibnamefont
  {Yamazaki}},\ }\bibfield  {title} {\enquote {\bibinfo {title} {{Crystal
  Melting and Toric Calabi-Yau Manifolds}},}\ }\href {\doibase
  10.1007/s00220-009-0836-y} {\bibfield  {journal} {\bibinfo  {journal}
  {Commun. Math. Phys.}\ }\textbf {\bibinfo {volume} {292}},\ \bibinfo {pages}
  {179--199} (\bibinfo {year} {2009})},\ \Eprint
  {http://arxiv.org/abs/0811.2801} {arXiv:0811.2801 [hep-th]} \BibitemShut
  {NoStop}%
\bibitem [{\citenamefont {Mozgovoy}\ and\ \citenamefont
  {Reineke}(2010)}]{MR2592501}%
  \BibitemOpen
  \bibfield  {author} {\bibinfo {author} {\bibfnamefont {S.}~\bibnamefont
  {Mozgovoy}}\ and\ \bibinfo {author} {\bibfnamefont {M.}~\bibnamefont
  {Reineke}},\ }\bibfield  {title} {\enquote {\bibinfo {title} {On the
  noncommutative {D}onaldson-{T}homas invariants arising from brane tilings},}\
  }\href {\doibase 10.1016/j.aim.2009.10.001} {\bibfield  {journal} {\bibinfo
  {journal} {Adv. Math.}\ }\textbf {\bibinfo {volume} {223}},\ \bibinfo {pages}
  {1521--1544} (\bibinfo {year} {2010})}\BibitemShut {NoStop}%
\bibitem [{\citenamefont {Yamazaki}(2011)}]{Yamazaki:2010fz}%
  \BibitemOpen
  \bibfield  {author} {\bibinfo {author} {\bibfnamefont {M.}~\bibnamefont
  {Yamazaki}},\ }\bibfield  {title} {\enquote {\bibinfo {title} {{Crystal
  Melting and Wall Crossing Phenomena}},}\ }\href {\doibase
  10.1142/S0217751X11051482} {\bibfield  {journal} {\bibinfo  {journal} {Int.
  J. Mod. Phys.}\ }\textbf {\bibinfo {volume} {A26}},\ \bibinfo {pages}
  {1097--1228} (\bibinfo {year} {2011})},\ \Eprint
  {http://arxiv.org/abs/1002.1709} {arXiv:1002.1709 [hep-th]} \BibitemShut
  {NoStop}%
\bibitem [{Note7()}]{Note7}%
  \BibitemOpen
  \bibinfo {note} {These formulas are inspired by their counterparts for
  $\protect \mathbb {C}^3$, see Refs.\ \protect \rev@citealpnum {feigin2012,
  Tsymbaliuk:2014fvq, Prochazka:2015deb}. Some more examples of toric
  Calabi-Yau manifolds also discussed e.g.\ in Refs.\ \protect \rev@citealpnum
  {Feigin1204,Feigin:2013fga,Gaberdiel:2017dbk,Rapcak:2018nsl,Rapcak:2020ueh}.}\BibitemShut
  {Stop}%
\bibitem [{Note8()}]{Note8}%
  \BibitemOpen
  \bibinfo {note} {In general we can play such a game for $\psi _0^{(a)}$ for
  each quiver vertex $a$.}\BibitemShut {Stop}%
\bibitem [{\citenamefont {Nagao}\ and\ \citenamefont
  {Nakajima}(2011)}]{MR2836398}%
  \BibitemOpen
  \bibfield  {author} {\bibinfo {author} {\bibfnamefont {K.}~\bibnamefont
  {Nagao}}\ and\ \bibinfo {author} {\bibfnamefont {H.}~\bibnamefont
  {Nakajima}},\ }\bibfield  {title} {\enquote {\bibinfo {title} {Counting
  invariant of perverse coherent sheaves and its wall-crossing},}\ }\href
  {\doibase 10.1093/imrn/rnq195} {\bibfield  {journal} {\bibinfo  {journal}
  {Int. Math. Res. Not. IMRN}\ }\textbf {\bibinfo {volume} {2011}},\ \bibinfo
  {pages} {3885--3938} (\bibinfo {year} {2011})},\ \Eprint
  {http://arxiv.org/abs/0809.2992} {arXiv:0809.2992 [math.AG]} \BibitemShut
  {NoStop}%
\bibitem [{\citenamefont {Jafferis}\ and\ \citenamefont
  {Moore}(2008)}]{Jafferis:2008uf}%
  \BibitemOpen
  \bibfield  {author} {\bibinfo {author} {\bibfnamefont {D.~L.}\ \bibnamefont
  {Jafferis}}\ and\ \bibinfo {author} {\bibfnamefont {G.~W.}\ \bibnamefont
  {Moore}},\ }\bibfield  {title} {\enquote {\bibinfo {title} {{Wall crossing in
  local Calabi Yau manifolds}},}\ }\href@noop {} {\  (\bibinfo {year}
  {2008})},\ \Eprint {http://arxiv.org/abs/0810.4909} {arXiv:0810.4909
  [hep-th]} \BibitemShut {NoStop}%
\bibitem [{\citenamefont {Chuang}\ and\ \citenamefont
  {Jafferis}(2009)}]{Chuang:2008aw}%
  \BibitemOpen
  \bibfield  {author} {\bibinfo {author} {\bibfnamefont {W.-y.}\ \bibnamefont
  {Chuang}}\ and\ \bibinfo {author} {\bibfnamefont {D.~L.}\ \bibnamefont
  {Jafferis}},\ }\bibfield  {title} {\enquote {\bibinfo {title} {{Wall Crossing
  of BPS States on the Conifold from Seiberg Duality and Pyramid
  Partitions}},}\ }\href {\doibase 10.1007/s00220-009-0832-2} {\bibfield
  {journal} {\bibinfo  {journal} {Commun. Math. Phys.}\ }\textbf {\bibinfo
  {volume} {292}},\ \bibinfo {pages} {285--301} (\bibinfo {year} {2009})},\
  \Eprint {http://arxiv.org/abs/0810.5072} {arXiv:0810.5072 [hep-th]}
  \BibitemShut {NoStop}%
\bibitem [{\citenamefont {Nagao}(2009)}]{Nagao:2009ky}%
  \BibitemOpen
  \bibfield  {author} {\bibinfo {author} {\bibfnamefont {K.}~\bibnamefont
  {Nagao}},\ }\bibfield  {title} {\enquote {\bibinfo {title} {{Refined open
  non-commutative Donaldson-Thomas invariants for small crepant
  resolutions}},}\ }\href@noop {} {\  (\bibinfo {year} {2009})},\ \Eprint
  {http://arxiv.org/abs/0907.3784} {arXiv:0907.3784 [math.AG]} \BibitemShut
  {NoStop}%
\bibitem [{\citenamefont {Nagao}\ and\ \citenamefont
  {Yamazaki}(2010)}]{Nagao:2009rq}%
  \BibitemOpen
  \bibfield  {author} {\bibinfo {author} {\bibfnamefont {K.}~\bibnamefont
  {Nagao}}\ and\ \bibinfo {author} {\bibfnamefont {M.}~\bibnamefont
  {Yamazaki}},\ }\bibfield  {title} {\enquote {\bibinfo {title} {{The
  Non-commutative Topological Vertex and Wall Crossing Phenomena}},}\ }\href
  {\doibase 10.4310/ATMP.2010.v14.n4.a3} {\bibfield  {journal} {\bibinfo
  {journal} {Adv. Theor. Math. Phys.}\ }\textbf {\bibinfo {volume} {14}},\
  \bibinfo {pages} {1147--1181} (\bibinfo {year} {2010})},\ \Eprint
  {http://arxiv.org/abs/0910.5479} {arXiv:0910.5479 [hep-th]} \BibitemShut
  {NoStop}%
\bibitem [{\citenamefont {Sulkowski}(2011)}]{Sulkowski:2010eg}%
  \BibitemOpen
  \bibfield  {author} {\bibinfo {author} {\bibfnamefont {P.}~\bibnamefont
  {Sulkowski}},\ }\bibfield  {title} {\enquote {\bibinfo {title}
  {{Wall-crossing, open BPS counting and matrix models}},}\ }\href {\doibase
  10.1007/JHEP03(2011)089} {\bibfield  {journal} {\bibinfo  {journal} {JHEP}\
  }\textbf {\bibinfo {volume} {03}},\ \bibinfo {pages} {089} (\bibinfo {year}
  {2011})},\ \bibinfo {note} {[Erratum: JHEP 04, 046 (2011)]},\ \Eprint
  {http://arxiv.org/abs/1011.5269} {arXiv:1011.5269 [hep-th]} \BibitemShut
  {NoStop}%
\bibitem [{\citenamefont {Gaiotto}\ and\ \citenamefont
  {Rap\v{c}\'ak}(2019)}]{Gaiotto:2017euk}%
  \BibitemOpen
  \bibfield  {author} {\bibinfo {author} {\bibfnamefont {D.}~\bibnamefont
  {Gaiotto}}\ and\ \bibinfo {author} {\bibfnamefont {M.}~\bibnamefont
  {Rap\v{c}\'ak}},\ }\bibfield  {title} {\enquote {\bibinfo {title} {{Vertex
  Algebras at the Corner}},}\ }\href {\doibase 10.1007/JHEP01(2019)160}
  {\bibfield  {journal} {\bibinfo  {journal} {JHEP}\ }\textbf {\bibinfo
  {volume} {01}},\ \bibinfo {pages} {160} (\bibinfo {year} {2019})},\ \Eprint
  {http://arxiv.org/abs/1703.00982} {arXiv:1703.00982 [hep-th]} \BibitemShut
  {NoStop}%
\bibitem [{Note9()}]{Note9}%
  \BibitemOpen
  \bibinfo {note} {In general, in addition to \protect \textup {\hbox
  {\mathsurround \z@ \protect \normalfont (\ignorespaces \ref
  {eq-OPE-toric}\unskip \@@italiccorr )}} there are extra relations---Serre
  relations---satisfied by the generators. When the Serre relations are
  included we have the reduced quiver Yangian.}\BibitemShut {Stop}%
\bibitem [{\citenamefont {Rapcak}(2021)}]{Rapcak:2021hdh}%
  \BibitemOpen
  \bibfield  {author} {\bibinfo {author} {\bibfnamefont {M.}~\bibnamefont
  {Rapcak}},\ }\bibfield  {title} {\enquote {\bibinfo {title} {{Branes, Quivers
  and BPS Algebras}},}\ }\href@noop {} {\  (\bibinfo {year} {2021})},\ \Eprint
  {http://arxiv.org/abs/2112.13878} {arXiv:2112.13878 [hep-th]} \BibitemShut
  {NoStop}%
\bibitem [{\citenamefont {Noshita}\ and\ \citenamefont
  {Watanabe}(2022{\natexlab{a}})}]{Noshita:2021ldl}%
  \BibitemOpen
  \bibfield  {author} {\bibinfo {author} {\bibfnamefont {G.}~\bibnamefont
  {Noshita}}\ and\ \bibinfo {author} {\bibfnamefont {A.}~\bibnamefont
  {Watanabe}},\ }\bibfield  {title} {\enquote {\bibinfo {title} {{A note on
  quiver quantum toroidal algebra}},}\ }\href {\doibase
  10.1007/JHEP05(2022)011} {\bibfield  {journal} {\bibinfo  {journal} {JHEP}\
  }\textbf {\bibinfo {volume} {05}},\ \bibinfo {pages} {011} (\bibinfo {year}
  {2022}{\natexlab{a}})},\ \Eprint {http://arxiv.org/abs/2108.07104}
  {arXiv:2108.07104 [hep-th]} \BibitemShut {NoStop}%
\bibitem [{\citenamefont {Galakhov}, \citenamefont {Li},\ and\ \citenamefont
  {Yamazaki}(2022)}]{Galakhov:2021vbo}%
  \BibitemOpen
  \bibfield  {author} {\bibinfo {author} {\bibfnamefont {D.}~\bibnamefont
  {Galakhov}}, \bibinfo {author} {\bibfnamefont {W.}~\bibnamefont {Li}}, \ and\
  \bibinfo {author} {\bibfnamefont {M.}~\bibnamefont {Yamazaki}},\ }\bibfield
  {title} {\enquote {\bibinfo {title} {{Toroidal and elliptic quiver BPS
  algebras and beyond}},}\ }\href {\doibase 10.1007/JHEP02(2022)024} {\bibfield
   {journal} {\bibinfo  {journal} {JHEP}\ }\textbf {\bibinfo {volume} {02}},\
  \bibinfo {pages} {024} (\bibinfo {year} {2022})},\ \Eprint
  {http://arxiv.org/abs/2108.10286} {arXiv:2108.10286 [hep-th]} \BibitemShut
  {NoStop}%
\bibitem [{\citenamefont {Noshita}\ and\ \citenamefont
  {Watanabe}(2022{\natexlab{b}})}]{Noshita:2021dgj}%
  \BibitemOpen
  \bibfield  {author} {\bibinfo {author} {\bibfnamefont {G.}~\bibnamefont
  {Noshita}}\ and\ \bibinfo {author} {\bibfnamefont {A.}~\bibnamefont
  {Watanabe}},\ }\bibfield  {title} {\enquote {\bibinfo {title} {{Shifted
  quiver quantum toroidal algebra and subcrystal representations}},}\ }\href
  {\doibase 10.1007/JHEP05(2022)122} {\bibfield  {journal} {\bibinfo  {journal}
  {JHEP}\ }\textbf {\bibinfo {volume} {05}},\ \bibinfo {pages} {122} (\bibinfo
  {year} {2022}{\natexlab{b}})},\ \Eprint {http://arxiv.org/abs/2109.02045}
  {arXiv:2109.02045 [hep-th]} \BibitemShut {NoStop}%
\bibitem [{\citenamefont {Kennaway}(2007)}]{Kennaway:2007tq}%
  \BibitemOpen
  \bibfield  {author} {\bibinfo {author} {\bibfnamefont {K.~D.}\ \bibnamefont
  {Kennaway}},\ }\bibfield  {title} {\enquote {\bibinfo {title} {{Brane
  Tilings}},}\ }\href {\doibase 10.1142/S0217751X07036877} {\bibfield
  {journal} {\bibinfo  {journal} {Int. J. Mod. Phys.}\ }\textbf {\bibinfo
  {volume} {A22}},\ \bibinfo {pages} {2977--3038} (\bibinfo {year} {2007})},\
  \Eprint {http://arxiv.org/abs/0706.1660} {arXiv:0706.1660 [hep-th]}
  \BibitemShut {NoStop}%
\bibitem [{\citenamefont {Yamazaki}(2008)}]{Yamazaki:2008bt}%
  \BibitemOpen
  \bibfield  {author} {\bibinfo {author} {\bibfnamefont {M.}~\bibnamefont
  {Yamazaki}},\ }\bibfield  {title} {\enquote {\bibinfo {title} {{Brane Tilings
  and Their Applications}},}\ }\href {\doibase 10.1002/prop.200810536}
  {\bibfield  {journal} {\bibinfo  {journal} {Fortsch. Phys.}\ }\textbf
  {\bibinfo {volume} {56}},\ \bibinfo {pages} {555--686} (\bibinfo {year}
  {2008})},\ \Eprint {http://arxiv.org/abs/0803.4474} {arXiv:0803.4474
  [hep-th]} \BibitemShut {NoStop}%
\bibitem [{\citenamefont {Bezerra}\ and\ \citenamefont
  {Mukhin}(2021)}]{Bezerra:2019dmp}%
  \BibitemOpen
  \bibfield  {author} {\bibinfo {author} {\bibfnamefont {L.}~\bibnamefont
  {Bezerra}}\ and\ \bibinfo {author} {\bibfnamefont {E.}~\bibnamefont
  {Mukhin}},\ }\bibfield  {title} {\enquote {\bibinfo {title} {Quantum toroidal
  algebra associated with {$\mathfrak{gl}_{m|n}$}},}\ }\href {\doibase
  10.1007/s10468-020-09959-9} {\bibfield  {journal} {\bibinfo  {journal}
  {Algebr. Represent. Theory}\ }\textbf {\bibinfo {volume} {24}},\ \bibinfo
  {pages} {541--564} (\bibinfo {year} {2021})}\BibitemShut {NoStop}%
\bibitem [{\citenamefont {Feigin}\ \emph {et~al.}(2012)\citenamefont {Feigin},
  \citenamefont {Jimbo}, \citenamefont {Miwa},\ and\ \citenamefont
  {Mukhin}}]{feigin2012}%
  \BibitemOpen
  \bibfield  {author} {\bibinfo {author} {\bibfnamefont {B.}~\bibnamefont
  {Feigin}}, \bibinfo {author} {\bibfnamefont {M.}~\bibnamefont {Jimbo}},
  \bibinfo {author} {\bibfnamefont {T.}~\bibnamefont {Miwa}}, \ and\ \bibinfo
  {author} {\bibfnamefont {E.}~\bibnamefont {Mukhin}},\ }\bibfield  {title}
  {\enquote {\bibinfo {title} {Quantum toroidal $\mathfrak{gl}_{1}$ algebra:
  Plane partitions},}\ }\href@noop {} {\bibfield  {journal} {\bibinfo
  {journal} {Kyoto J. Math.}\ }\textbf {\bibinfo {volume} {52}},\ \bibinfo
  {pages} {621--659} (\bibinfo {year} {2012})}\BibitemShut {NoStop}%
\bibitem [{\citenamefont {Tsymbaliuk}(2017)}]{Tsymbaliuk:2014fvq}%
  \BibitemOpen
  \bibfield  {author} {\bibinfo {author} {\bibfnamefont {A.}~\bibnamefont
  {Tsymbaliuk}},\ }\bibfield  {title} {\enquote {\bibinfo {title} {{The affine
  Yangian of $\mathfrak{gl}_1$ revisited}},}\ }\href {\doibase
  10.1016/j.aim.2016.08.041} {\bibfield  {journal} {\bibinfo  {journal} {Adv.
  Math.}\ }\textbf {\bibinfo {volume} {304}},\ \bibinfo {pages} {583--645}
  (\bibinfo {year} {2017})},\ \Eprint {http://arxiv.org/abs/1404.5240}
  {arXiv:1404.5240 [math.RT]} \BibitemShut {NoStop}%
\bibitem [{\citenamefont {Proch\'azka}(2016)}]{Prochazka:2015deb}%
  \BibitemOpen
  \bibfield  {author} {\bibinfo {author} {\bibfnamefont {T.}~\bibnamefont
  {Proch\'azka}},\ }\bibfield  {title} {\enquote {\bibinfo {title} {{$
  \mathcal{W} $ -symmetry, topological vertex and affine Yangian}},}\ }\href
  {\doibase 10.1007/JHEP10(2016)077} {\bibfield  {journal} {\bibinfo  {journal}
  {JHEP}\ }\textbf {\bibinfo {volume} {10}},\ \bibinfo {pages} {077} (\bibinfo
  {year} {2016})},\ \Eprint {http://arxiv.org/abs/1512.07178} {arXiv:1512.07178
  [hep-th]} \BibitemShut {NoStop}%
\bibitem [{\citenamefont {{Feigin}}\ \emph {et~al.}()\citenamefont {{Feigin}},
  \citenamefont {{Jimbo}}, \citenamefont {{Miwa}},\ and\ \citenamefont
  {{Mukhin}}}]{Feigin1204}%
  \BibitemOpen
  \bibfield  {author} {\bibinfo {author} {\bibfnamefont {B.}~\bibnamefont
  {{Feigin}}}, \bibinfo {author} {\bibfnamefont {M.}~\bibnamefont {{Jimbo}}},
  \bibinfo {author} {\bibfnamefont {T.}~\bibnamefont {{Miwa}}}, \ and\ \bibinfo
  {author} {\bibfnamefont {E.}~\bibnamefont {{Mukhin}}},\ }\bibfield  {title}
  {\enquote {\bibinfo {title} {{Representations of quantum toroidal
  $\mathfrak{gl}_n$}},}\ }\href@noop {} {\ }\Eprint
  {http://arxiv.org/abs/1204.5378} {arXiv:1204.5378 [math.QA]} \BibitemShut
  {NoStop}%
\bibitem [{\citenamefont {Feigin}\ \emph {et~al.}(2016)\citenamefont {Feigin},
  \citenamefont {Jimbo}, \citenamefont {Miwa},\ and\ \citenamefont
  {Mukhin}}]{Feigin:2013fga}%
  \BibitemOpen
  \bibfield  {author} {\bibinfo {author} {\bibfnamefont {B.}~\bibnamefont
  {Feigin}}, \bibinfo {author} {\bibfnamefont {M.}~\bibnamefont {Jimbo}},
  \bibinfo {author} {\bibfnamefont {T.}~\bibnamefont {Miwa}}, \ and\ \bibinfo
  {author} {\bibfnamefont {E.}~\bibnamefont {Mukhin}},\ }\bibfield  {title}
  {\enquote {\bibinfo {title} {{Branching rules for quantum toroidal
  gl$_n$}},}\ }\href {\doibase 10.1016/j.aim.2016.03.019} {\bibfield  {journal}
  {\bibinfo  {journal} {Adv. Math.}\ }\textbf {\bibinfo {volume} {300}},\
  \bibinfo {pages} {229--274} (\bibinfo {year} {2016})},\ \Eprint
  {http://arxiv.org/abs/1309.2147} {arXiv:1309.2147 [math.QA]} \BibitemShut
  {NoStop}%
\bibitem [{\citenamefont {Gaberdiel}\ \emph {et~al.}(2017)\citenamefont
  {Gaberdiel}, \citenamefont {Gopakumar}, \citenamefont {Li},\ and\
  \citenamefont {Peng}}]{Gaberdiel:2017dbk}%
  \BibitemOpen
  \bibfield  {author} {\bibinfo {author} {\bibfnamefont {M.~R.}\ \bibnamefont
  {Gaberdiel}}, \bibinfo {author} {\bibfnamefont {R.}~\bibnamefont
  {Gopakumar}}, \bibinfo {author} {\bibfnamefont {W.}~\bibnamefont {Li}}, \
  and\ \bibinfo {author} {\bibfnamefont {C.}~\bibnamefont {Peng}},\ }\bibfield
  {title} {\enquote {\bibinfo {title} {{Higher Spins and Yangian
  Symmetries}},}\ }\href {\doibase 10.1007/JHEP04(2017)152} {\bibfield
  {journal} {\bibinfo  {journal} {JHEP}\ }\textbf {\bibinfo {volume} {04}},\
  \bibinfo {pages} {152} (\bibinfo {year} {2017})},\ \Eprint
  {http://arxiv.org/abs/1702.05100} {arXiv:1702.05100 [hep-th]} \BibitemShut
  {NoStop}%
\bibitem [{\citenamefont {Rapcak}\ \emph {et~al.}(2019)\citenamefont {Rapcak},
  \citenamefont {Soibelman}, \citenamefont {Yang},\ and\ \citenamefont
  {Zhao}}]{Rapcak:2018nsl}%
  \BibitemOpen
  \bibfield  {author} {\bibinfo {author} {\bibfnamefont {M.}~\bibnamefont
  {Rapcak}}, \bibinfo {author} {\bibfnamefont {Y.}~\bibnamefont {Soibelman}},
  \bibinfo {author} {\bibfnamefont {Y.}~\bibnamefont {Yang}}, \ and\ \bibinfo
  {author} {\bibfnamefont {G.}~\bibnamefont {Zhao}},\ }\bibfield  {title}
  {\enquote {\bibinfo {title} {{Cohomological Hall algebras, vertex algebras
  and instantons}},}\ }\href {\doibase 10.1007/s00220-019-03575-5} {\bibfield
  {journal} {\bibinfo  {journal} {Commun. Math. Phys.}\ }\textbf {\bibinfo
  {volume} {376}},\ \bibinfo {pages} {1803--1873} (\bibinfo {year} {2019})},\
  \Eprint {http://arxiv.org/abs/1810.10402} {arXiv:1810.10402 [math.QA]}
  \BibitemShut {NoStop}%
\bibitem [{\citenamefont {Rapcak}\ \emph {et~al.}(2020)\citenamefont {Rapcak},
  \citenamefont {Soibelman}, \citenamefont {Yang},\ and\ \citenamefont
  {Zhao}}]{Rapcak:2020ueh}%
  \BibitemOpen
  \bibfield  {author} {\bibinfo {author} {\bibfnamefont {M.}~\bibnamefont
  {Rapcak}}, \bibinfo {author} {\bibfnamefont {Y.}~\bibnamefont {Soibelman}},
  \bibinfo {author} {\bibfnamefont {Y.}~\bibnamefont {Yang}}, \ and\ \bibinfo
  {author} {\bibfnamefont {G.}~\bibnamefont {Zhao}},\ }\bibfield  {title}
  {\enquote {\bibinfo {title} {{Cohomological Hall algebras and perverse
  coherent sheaves on toric Calabi-Yau 3-folds}},}\ }\href@noop {} {\
  (\bibinfo {year} {2020})},\ \Eprint {http://arxiv.org/abs/2007.13365}
  {arXiv:2007.13365 [math.QA]} \BibitemShut {NoStop}%
\end{thebibliography}%

\end{document}